\documentstyle[prd,aps]{revtex}

\def\be{\begin{equation}} 
\def\ee{\end{equation}} 
\def\bea{\begin{eqnarray}} 
\def\eea{\end{eqnarray}} 
\begin{document}
\draft
\title{\bf Defect-Antidefect Pair Production via Field Oscillations}
\author{Sanatan Digal, Supratim Sengupta and Ajit M. 
Srivastava \footnote{E-mail :
digal,supratim,ajit@iopb.ernet.in}}
\address{Institute of Physics\\
Sachivalaya Marg, Bhubaneswar--751005, INDIA}
\twocolumn[
\maketitle
\widetext
\parshape=1 0.75in 5.5in
\begin{abstract}
 We show that the mechanism of vortex-antivortex pair production via 
field oscillations, earlier proposed by two of us for systems with 
first order transitions in presence of explicit symmetry breaking, 
is in fact very generally applicable. We further argue that this 
mechanism applies to all sorts of topological defects (e.g. strings, 
monopoles, textures) and for second order transitions as well. We
also show this explicitly  by numerically simulating production of
vortex-antivortex pairs by decay of bubble walls for a U(1) global
theory in the absence of any explicit symmetry breaking.
\end{abstract}
\vskip 0.125 in
\parshape=1 0.75in 5.5in
]
\narrowtext
 
 Recently it has been suggested that the process of formation
of topological defects in phase transitions may be much more 
complex than realized earlier. For example in \cite{dgl} it 
was demonstrated (for first order transitions in 2+1 dimensions) 
that for the case when a U(1) global symmetry
is spontaneously as well as explicitly broken then the vortex
production gets dominated by a new process. In this process,
oscillations of the order parameter field in the coalesced 
portions of bubble walls lead to the decay of the wall in a large 
number of vortex-antivortex pairs (e.g. 5 pairs in a single 
two bubble collision). 

 However, systems with explicit symmetry breaking form a very 
special subclass of all the physical systems where topological 
defects can form in phase transitions \cite{shlrd}. 
Therefore, it is very important
to find out whether this mechanism can also play any role in defect 
formation in other systems as well. We study this issue in this 
paper. We analyze the basic physical picture underlying this 
mechanism and show that it is applicable even in the absence of
explicit symmetry breaking and also for the formation of
all sorts of topological defects, for example, strings, monopoles,
and textures. Furthermore, the field oscillations required for this
mechanism, which naturally arise by decay of bubble walls for first 
order transitions (as in \cite{dgl}), can also arise for second
order transitions if one considers quench from sufficiently high 
temperature. This mechanism of defect formation is, therefore, 
completely generally applicable. 

  We start by presenting the physical picture of this mechanism. 
Consider first the formation of U(1) global vortices in 2+1
dimensions with the order parameter being a complex scalar field
$\Phi$. Let us assume that the phase $\theta$ of $\Phi$ varies 
in a region of space as shown in Fig.1a, changing from  some value 
$\alpha$ to some different value $\beta$ uniformly 
as we go from bottom to top. At this stage there is no vortex 
present in this region.  Now assume that $\Phi$ undergoes 
oscillation, passing once through $\Phi = 0$, in a small region 
in the center enclosed by the dotted loop, see Fig.1b. It is 
clear that when $\Phi$ passes through 0, it amounts to discontinuous 
change in $\theta$ by $\pi$, as shown in Fig.1b.  We call it the 
flipping of $\Phi$. [For simplicity, we take $\theta$ to be uniform 
inside the flipped region.] Now consider the variation of $\theta$ 
along the closed path AOBCD (shown by the solid curve in 
Fig.1b).   On moving along this solid curve, we take $\theta$ to 
vary smoothly (as shown by the dotted arrows) as we cross the 
dotted curve. [Our results do not depend on whether one 
goes along the shortest path on the vacuum manifold S$^1$, or 
along the longer path, as we cross the dotted curve.  In the 
later case, the vortex and the antivortex get interchanged.] 
It is then easy to see that $\theta$ winds by $2\pi$ as we go 
around this closed path implying that a vortex has formed inside 
it. Similarly, we can conclude that an antivortex has formed in 
the other half of this region.

 Clearly, these arguments do not require the presence of explicit 
symmetry breaking for vortex-antivortex pair production. Another 
way to see the pair formation, which is also easier to generalize 
for other defects, is the following. Consider the variation of
$\theta$ along path AOBCD in Fig.1b, first before the flipping
of $\Phi$ in the dotted region. Clearly $\theta$ traces an arc
{\it P} on the vacuum manifold $S^1$ (clockwise from $\beta$ to 
$\alpha$) as we traverse the path AOB and then traces this arc 
backwards (from $\alpha$ to $\beta$) as we traverse the segment 
BCDA continuing along the solid path. This is a contractible loop 
on S$^1$. After flipping of $\Phi$ in the dotted region, a portion 
in the middle of the arc {\it P} on $S^1$ moves to the opposite
side on this $S^1$. [Our results do not depend on whether the
orientation of this portion also gets reversed in this flipping,
or not.] 

 If the mid point of the arc {\it P} was at $\theta = \gamma$ 
initially (corresponding to point O in Fig.1b), then $\gamma$ 
changes to $\gamma + \pi$ by this flip. As one goes from A to 
O, one moves from $\theta = \beta$ on $S^1$, initially along 
the arc {\it P}, but then along the circle either anti-clockwise,
or clockwise, in order to reach the value $\gamma + \pi$. As 
we continue from O to B, we now have to travel along the other 
side on $S^1$ (i.e. anti-clockwise or clockwise respectively) 
due to symmetry of the situation, now starting at $\gamma + \pi$ 
and ending at $\theta = \alpha$.  As we continue moving along 
the solid curve from B to C to D and back to A, we will trace 
the arc {\it P} backward (i.e. from $\theta = \alpha$ to $\gamma$ 
to $\beta$).  One can immediately see that net variation in 
$\theta$ is by 2$\pi$ for the first case, when $\theta$ varies
clockwise along the segment A to O (this is the case shown 
in Fig.1b and may be preferred due to energetic reasons), and 
by $-2\pi$ in the other case. Thus a vortex (antivortex) has 
formed inside the region enclosed by the solid curve. Clearly 
an antivortex (vortex) will form in the left portion of the 
Fig.1b. We should mention here that successive passage of $\Phi$
through zero will create further windings. At the same time,
density waves generated during oscillations leads to further
separation of a vortex and an antivortex created earlier.

  It is simple to generalize this picture for the formation
of other defects. Consider monopole formation in 3+1 dimensions
with the order parameter space being a two sphere $S^2$. Assume
that the order parameter is varying in a region of space in
a such a way that its variation in a disk like region in space 
covers a small patch on $S^2$ at its north pole (by convention). 
Further assume (for simplicity) that the order parameter does not 
vary much in direction normal to the disk, thus forming a cylindrical
segment like region. Assume now that the order parameter 
flips (as it passes through zero once, while oscillating),
in a small volume near the center of this cylindrical
segment (midway along its length) meaning that in that region, 
the order parameter now take values on a (smaller) patch at 
the south pole of $S^2$. Consider a closed surface in the physical 
space starting from the center of the cylindrical region and 
enclosing one half of the flipped region completely. Variation
of order parameter on this surface starts from the south pole of
$S^2$ (corresponding to the center of the cylindrical region), 
reaches the boundary of patch at the north pole (covering the 
surface of $S^2$) and then continues to cover the patch at the north 
pole, eventually ending at the north pole. This amounts to one full 
winding on $S^2$ and hence a monopole being enclosed by the
closed surface in the physical space. As the order parameter has
trivial winding  around the full cylindrical portion (due to 
the flipping being localized) the other half of the cylinder must
get an antimonopole produced. Again, in this case also, by changing
the manner in which one interpolates between the south pole
and the boundary of the patch at the north pole of $S^2$,
one can see that one still gets a monopole-antimonopole pair.

 Formation of textures (Skyrmions)  is even easier to understand. 
First take a 2+1 dimensional example where textures (usually called
as baby Skyrmions) arise due to compactifying a portion of the 
physical space to $S^2$ by constant boundary conditions and then 
considering nontrivial mappings to the vacuum manifold $S^2$. 
Assume that the variation of the order parameter on a small 
disk {\it D} in the physical space covers a small patch at the 
north pole of the vacuum manifold $S^2$. Consider flipping of the 
order parameter near the center of this disk leading to order 
parameter in that region forming a patch at the 
south pole (just like the monopole case above).
We can immediately see that the variation of order parameter from 
the center of the disk {\it D} to the boundary of {\it D} amounts
to covering $S^2$ starting from its south pole, all the way up to
the boundary of the patch at the north pole. Note that this does
not complete full winding. However, as shown in \cite{skrm},
full winding textures are any way extremely suppressed in phase 
transitions. What one expects to get is the formation of partial 
winding textures. One can then argue that textures with windings 
greater than about 0.5 will evolve to have full windings. Thus for
small patches near the north pole (i.e. small spatial variations 
in the order parameter) one should easily form textures with
windings almost close to 1 (or -1) eventually leading to full
winding textures. 

 Extension to textures (Skyrmions) in 3+1 dimensions is almost obvious.
One considers a ball {\it B} in the physical space with the order parameter
in this region corresponding to a small neighborhood of the north pole
of the vacuum manifold (which is a three sphere $S^3$ now). Flipping
of the order parameter at the center of the ball {\it B} will lead
to the order parameter covering a portion of the south pole of $S^3$
in that region. Then, as we move from the center of the ball {\it B}
to its boundary in the physical space, it will lead to almost full
covering of $S^3$ (starting from its south pole, up to the boundary
of the patch at the north pole). Again from the arguments of 
\cite{skrm}, it will evolve in full winding textures for small
patches at the north pole.

 This case of textures (Skyrmions) in 3+1 dimensions is extremely 
important as in the context of chiral models, baryons are supposed 
to be Skyrmion. Thus these processes may play role in the formation of
baryons in the heavy ion collisions (see, \cite{kpst}). [Though, 
within this framework one has to worry about how to implement baryon 
number conservation as here one can always create a single Skyrmion, 
in contrast to string or monopoles which are always pair produced.]

 So far we have only used the flipping of $\Phi$ but never mentioned
how it is expected to happen. Large oscillations of the order 
parameter field naturally arise for first order transitions in the 
coalesced portions of bubble walls \cite{dgl,ajt}. However, it is 
easy to argue that such oscillations can arise even in second 
order transition. Consider the case of a second order transition where 
the initial state is at a very high temperature amounting to large 
fluctuations in the amplitude of the field (of course with the vacuum 
expectation value equal to zero). Consider the transition carried out 
by a sudden quench to the low temperature phase with some symmetry being 
spontaneously broken. As the change in the shape of the effective
potential is sudden, the field configuration in many regions will
still have very large magnitudes. While settling down to the
non-zero vacuum expectation value, the field will undergo
oscillations. Clearly for sufficiently large initial
magnitude of the field (i.e. for high enough temperatures)
the oscillations will be energetic enough to take the field
all the way across the potential hill amounting to
the flipping of the order parameter in that region. This qualitative
argument shows that even in the case of second order transition, 
this mechanism should be applicable, provided one considers quenching 
from very high temperatures. It will be very interesting to verify this
explicitly in numerical simulations.

  To summarize our discussion so far, we have argued that this 
mechanism of defect-antidefect production via field oscillations 
is indeed very general and does not necessarily require the presence 
of explicit symmetry breaking. Moreover, it applies to all sorts 
of topological defects and to first order as well as second order 
transition. It is also important to note that in whatever we have said 
above it was not relevant whether we are dealing with a gauge theory 
or with a global theory. One only needed information about the order
parameter field and the fact that it flips in a small region. Thus
this mechanism should be applicable to gauge theories as well as
global theories. Of course the relative importance of this mechanism
will certainly depend on the specific system being considered, 
especially because it depends on the detailed dynamics of the
order parameter field. 

We may mention here that, in view of our results in
this paper, one may understand the formation of many vortex-antivortex
pairs with strongly overlapping configurations in \cite{ajt} as 
being due to this mechanism (as was speculated in \cite{dgl})
even though there was no explicit symmetry breaking present in 
\cite{ajt}. Copeland and Saffin have recently shown that the geodesic
rule (the assumption of shortest variation of the order parameter on
the vacuum manifold in between two adjacent domains) can get violated 
and defects can form in the coalesced region of bubble walls in
Abelian Higgs model\cite{cs}. In fact it has been
emphasized in \cite{cs} that due to the presence of gauge fields,
the equations of motion involve a force term for $\theta$ leading to
nontrivial dynamics of $\theta$. In this sense the mechanism of
defect production in \cite{cs} also seems to be similar to the
one discussed in \cite{dgl}.  

  We now present our results of numerical 
simulations of a particular case, that is, formation of global U(1) 
vortices in 2+1 dimensions for a first order transition. We show that 
well formed vortex-antivortex pairs are produced in bubble collisions
via this mechanism even though there is no explicit symmetry breaking
present here.  We study vortex-antivortex pair production in  2+1 
dimensions in a system described by the Lagrangian

\begin{equation}
{\it L} = {1 \over 2} \partial_{\mu} \Phi^{\dag} \partial^{\mu} \Phi
- {1 \over 4} \phi^2 (\phi - 1)^2 + \epsilon \phi^3 
\end{equation}
 
 This Lagrangian is expressed in terms of the dimension less field $\Phi$
and appropriately scaled coordinates, with $\phi$ and $\theta$ being 
the magnitude and phase of the complex field $\Phi$. At zero              
temperature the phase transition proceeds by nucleation of bubbles of true 
vacuum in the background of false vacuum via quantum tunneling.
The bubble-profile (the bounce solution) is obtained by solving the 
Eucledian field equation

\be
{d^2 \phi \over dr^2} + {2 \over r}{d \phi \over dr} -V^{\prime}(\phi)=0
\ee

\noindent where $V(\phi)$ is the effective potential in Eq.(1) and r 
is the radial coordinate in Eucledian space. In Minkowski space the 
initial bubble profile is obtained by setting t=0
in the bounce solution after it has been analytically continued to  
Minkowski space. The subsequent evolution of the bubble is 
governed by the following classical field equations in Minkowski space.

\be
\Box \Phi_i =-{\partial V(\Phi) \over \partial \Phi_i}, ~i=1,2
\ee

\noindent where $\Phi=\Phi_1 + i\Phi_2$, and time derivatives of the 
fields are set equal to zero at t=0. 

  From the form of $V(\phi)$ we can easily understand the nature 
of $\Phi$ oscillations when bubbles coalesce. When two bubbles meet, 
$\Phi$ in the coalesced portion of their walls does not immediately
decay to its vacuum expectation value and keeps oscillating about it
for some time. If $\Phi$ oscillations 
are sufficiently energetic in that region, then 
$\Phi$ may be able to climb the potential hill and overshoot the
value $\Phi = 0$. From the arguments given above (Fig.1), this 
then should lead to a pair formation.  It is clear that for the 
vortex and the antivortex to be well separated and well formed, 
it is important that the value of $\phi$ should not be too close 
to zero in between them. This implies that $\Phi$ while overshooting 
the value zero, should be able to climb the potential hill in the 
same direction and roll down to the other side of $V(\phi)$. 
In the following, we will show that such energetic $\Phi$ 
oscillations in the coalesced region are easily created
in a three bubble collision which then lead to the decay of
wall in a well separated vortex-antivortex pair. 

  The numerical techniques adapted in this investigation have been
described in detail in \cite{ajt}. Bubble nucleation consists of
replacing a region of false vacuum by the bubble profile with 
$\theta$ being uniform inside each bubble. Initial configuration is
taken to be consisting of three half bubbles located at lattice
boundaries. This initial field configuration 
is evolved by using a stabilized 
leapfrog algorithm of second order accuracy in both space and time
using free boundary conditions. The physical size of the lattice 
was taken to be 57.6 x 57.6 with $\bigtriangleup x = 0.048$ and 
$\bigtriangleup t = \bigtriangleup x/\sqrt{2}$, which satisfied the 
Courant stability criterion. Simulations were carried out on 
a HP-735 workstation at the Institute of Physics,Bhubaneswar.

Initial bubble separation was chosen to be large to ensure that the 
walls acquire sufficient energy, by the time the bubbles collide. 
Fig.2a shows the phase plot of the initial field configuration
(consisting of half bubbles at lattice boundaries). The phases
inside the two bubbles at one end of the lattice are $18^{\circ}$ and     
$342^{\circ}$ whereas the phase inside the bubble at the other end is
$9^{\circ}$.The bubbles on the left are placed symmetrically with 
respect to the third bubble.

While there is nothing special about this choice of phases, nor about 
the choice of center coordinates of the bubbles, we would nevertheless 
like to emphasize that it is important in the above example that we 
choose the phases such that the difference in phase inside one bubble 
and the average value of phases inside the other two bubbles was small. 
[In fact, even for the cases when phase difference is small between any 
two bubbles, we got reasonably well separated vortices.] This ensures that
a major fraction of the energy stored in the bubble walls is available
for $\phi$ oscillations and only a small fraction is utilized in
smoothening out the phase gradient in the coalesced portion of the 
bubbles. [For large phase differences, field oscillations are not
energetic enough to take the field to the other side of the
effective potential.]

However, even if phase difference is very small, the dissipation of 
the wall energy occurs over a large region if $\theta$ does not have 
much spatial variation in that region. This will result in the flipping 
of phase over the entire width of the coalesced region resulting in
$\phi$ oscillations which are not energetic enough. So, even 
though the magnitude of the scalar field overshoots the value zero
while oscillating (leading to a pair being produced), it remains 
in the valley of the effective potential around $\phi=0$ and does not 
roll down to the other side. In such a situation, one does not get
a well separated vortex-antivortex pair. [This is what seems to
happen if one tries to get sufficiently energetic $\phi$ oscillations
in a two bubble collision.]
It is useful here to contrast this situation with the one discussed
in \cite{dgl} involving explicit symmetry breaking. In that case,
energy of $\phi$ oscillations arose not only from
the kinetic energy\ of the bubble walls (as in the present case) 
but also from the potential energy stored in the walls due to the tilt in
the effective potential (for non-zero phase inside the bubble). That
is why many more (well separated) pairs formed in the case
discussed in \cite{dgl}.

Fig.2b shows the plot of $\Phi$ at $t = 42.4$. Variation of $\theta$ in 
the coalesced portion, combined with the flipping of the phase in 
a small region has led to the formation of a vortex-antivortex pair. 
Oscillation and subsequent decay of the wall in between the 
vortex-antivortex pair gives rise to density waves which leads to the 
separation of the vortex-antivortex pair. We see that 
at this stage, the vortex-antivortex pair 
are not well formed and have highly distorted configurations 
as seen in the surface plot of $-\phi$ in Fig.2c.  Fig.2d shows 
the surface plot of $-\phi$ at $t = 49.2$. The vortex and
the antivortex are very well separated by now and have acquired
roughly cylindrically symmetric configurations. During subsequent
evolution,the vortex and the antivortex come back together due
to the attractive force between them, and finally annihilate. 

 We mention here that initial phases in the bubbles (Fig.2a) were
such that no defect could have formed inside that region via the Kibble 
mechanism \cite{kbl} (as the variation of $\theta$, using the 
geodesic rule, along any closed loop within that region corresponds 
to a shrinkable loop in the order parameter space). The fact that
vortex-antivortex pair is forming inside this region indicates a
breakdown of the geodesic rule.  In fact, it has been shown by
Copeland and Saffin recently \cite{cs} that in a gauge theory,
violation of geodesic rule can occur due to oscillations in the
coalesced region of bubbles. We can understand it more generally 
as we know that  flipping of $\Phi$ due to oscillations always causes 
$\theta$ to discontinuously  change by $\pi$. In such a situation 
there seems no reason to expect that the geodesic rule should still 
be applicable. Important point to realize is that this situation
can arise not only in a gauge theory but also in a global theory.

 We conclude by emphasizing the main aspect of our results.
The fact that we find the mechanism to be completely generally
applicable is extremely important as now it may be possible to 
experimentally test it in a variety of condensed matter systems. 
Especially important will be to investigate the role of this 
mechanism in the formation of strings in superfluid 
He$^4$ involving quench from large temperatures (as it is
a second order transition), or  in He$^3$-A to He$^3$-B 
transition (which is a first order transition). Due to being
superfluids, field oscillations will not be damped in these
cases, which is very crucial for this mechanism. [In this context
we should mention that when explicit symmetry breaking is also 
present then even in the presence of damping one always gets at 
least on vortex-antivortex pair produced via this mechanism 
\cite{sprtm}.] Our numerical results in this paper concretely 
show that well separated 
vortex-antivortex pairs can form in first order transitions.
It is easy to see that for 3+1 dimensions these results imply
the formation of a well defined string loop whose size must be 
smaller than the size of the coalescing bubbles \cite{dgl}. 
String loops of such small sizes can not be explained
in terms of Kibble mechanism. Thus for example, for superfluid 
helium if one ever observes such small loops 
experimentally then one will have to conclude that those loops 
formed due to this mechanism. [By choosing quench to small enough 
temperature, one can completely suppress thermal production of 
loops.] Such tests are very important (just as tests for the 
Kibble mechanism, see \cite{expt}) in providing a solid foundation
for the theories of defect formation which can then be applied
to situations which are otherwise experimentally inaccessible.
For example, this mechanism (just as the thermal production,
or the Kibble mechanism) should also contribute to the formation
of a variety of defects in the early Universe \cite{shlrd}. Our 
results also show that many vortex-antivortex pairs should form 
with overlapping configurations. This implies that for 3+1 
dimensional case (say that of He$^3$-A to He$^3$-B transition), 
many string loops with very small sizes should form. This 
mechanism can,  thus, affect the manner in which the transition 
to the low temperature phase is completed. 


\vskip .3in
\centerline {\bf FIGURE CAPTIONS}
\vskip .1in

Fig.1: (a) A region of space with $\theta$ varying uniformly
from $\alpha$ at the bottom to some value $\beta$ at the top.
(b) Flipping of $\Phi$ in the center (enclosed by the dotted
loop) has changed $\theta = \gamma$ to $\theta = \gamma + \pi$ 
resulting in a pair production.

Fig.2: (a) Plot of $\Phi$ for the initial configuration. Length of
a vector is proportional to $\phi$ while its orientation from the
x axis corresponds to $\theta$. (b) A small portion of the coalesced 
region at $t = 42.4$. Flipping of $\Phi$ in this region has resulted 
in the production of a vortex and an antivortex. (c) Surface
plot of $-\phi$ at $t = 42.4$ showing that vortices are not well
formed yet. (d) Surface plot at $t = 49.2$ showing the profiles of
well formed, and well separated vortex and antivortex. 

\begin{references}

\bibitem{dgl} S. Digal and A.M. Srivastava, Phys. Rev. Lett.
{\bf 76}, 583 (1996).

\bibitem{shlrd} For a review see, A. Vilenkin and E.P.S. Shellard,
``Cosmic strings and other topological defects", (Cambridge
University Press, Cambridge, 1994).

\bibitem{skrm} A. M. Srivastava, Phys. Rev. D {\bf 43}, 1047 (1991);
J. Borrill, E. J. Copeland and A. R. Liddle, Phys. Lett. B {\bf 258}, 
310 (1991).

\bibitem{kpst} J. I. Kapusta and A. M. Srivastava, Phys. Rev. D
{\bf 52}, 2977 (1995).

\bibitem{ajt} A.M. Srivastava, Phys. Rev. {\bf D45}, R3304 (1992);
{\bf D46}, 1353 (1992); S. Chakravarty and A.M. Srivastava,
Nucl. Phys. {\bf B406}, 795 (1993).

\bibitem{cs} E.J. Copeland and P.M. Saffin, preprint, SUSX-TH-96-006.

\bibitem{kbl} T.W.B. Kibble, J. Phys. {\bf A9}, 1387 (1976).
 
\bibitem{sprtm} S. Digal, S. Sengupta and A.M. Srivastava, in 
preparation.

\bibitem{expt} For a review see, W.H. Zurek, preprint,
Los Alamos National Laboratory (1996).

\end{references}
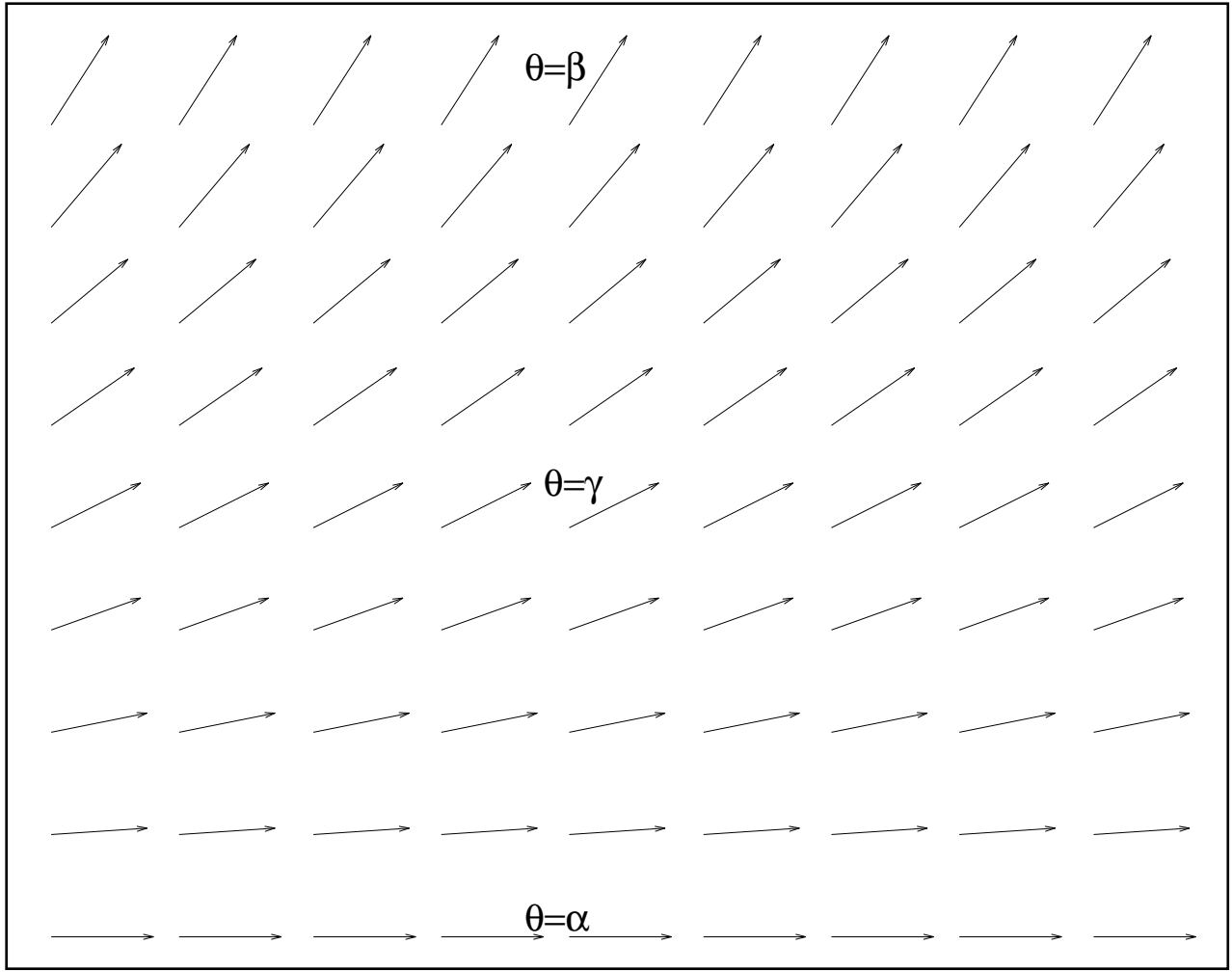
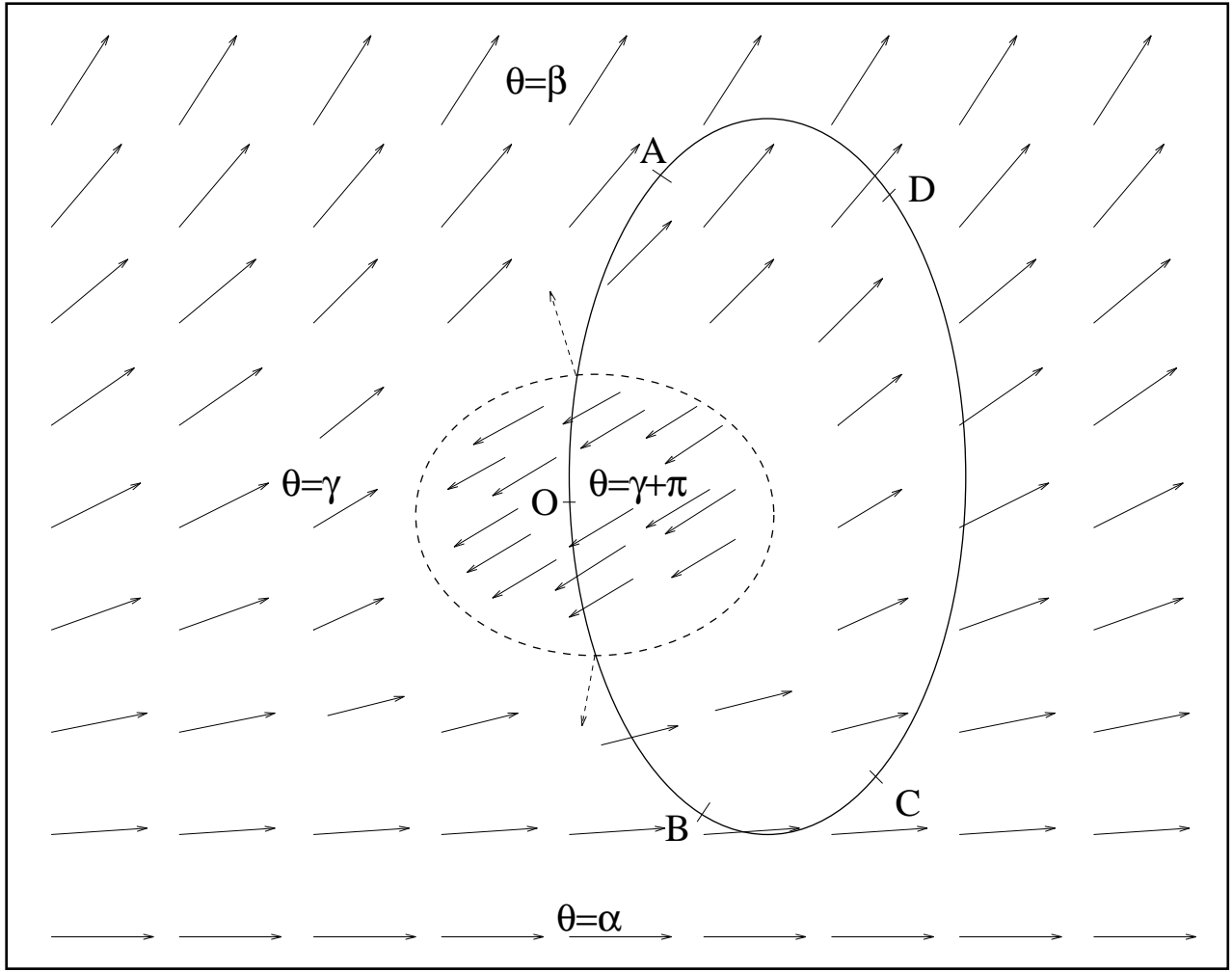
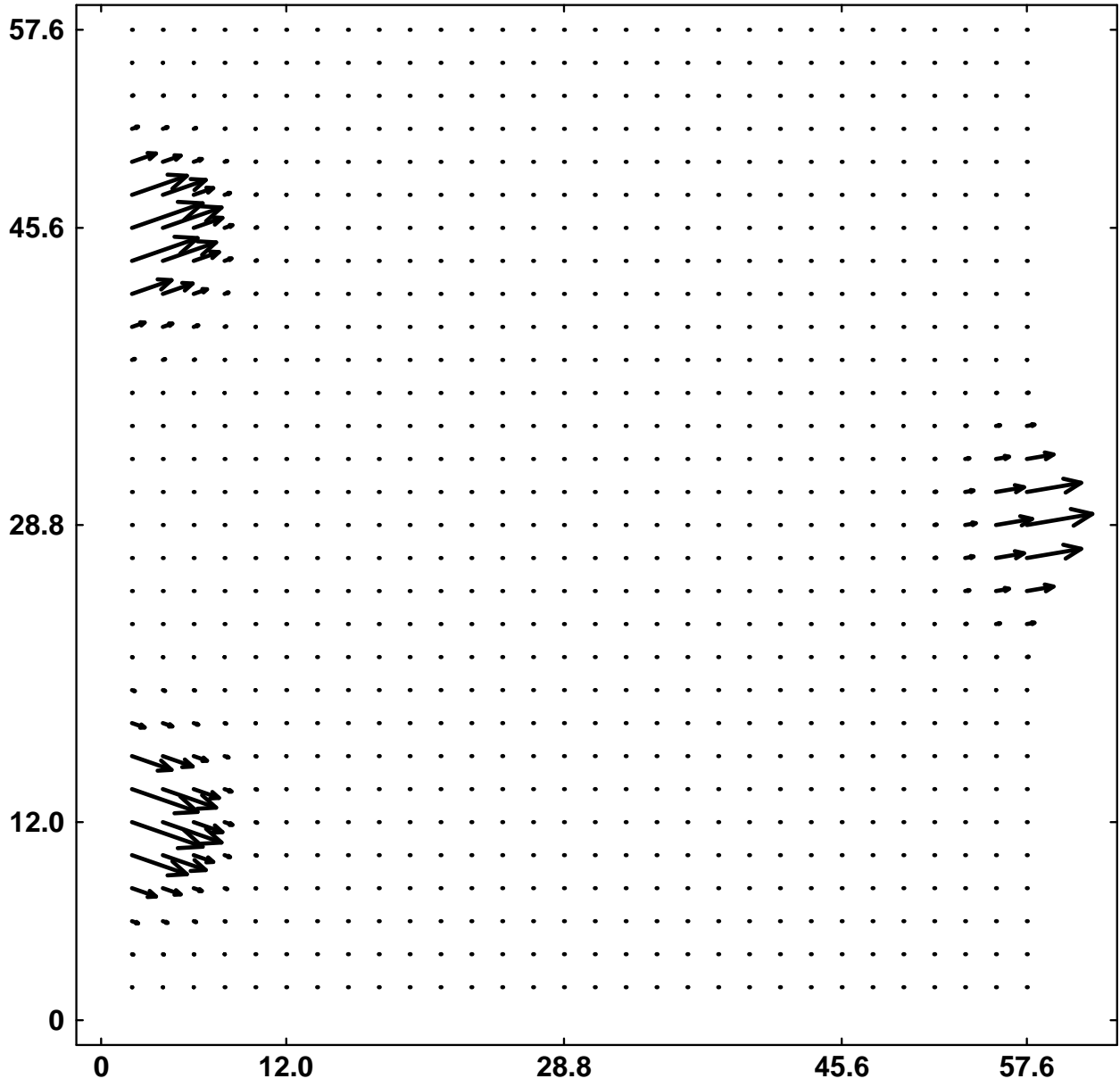
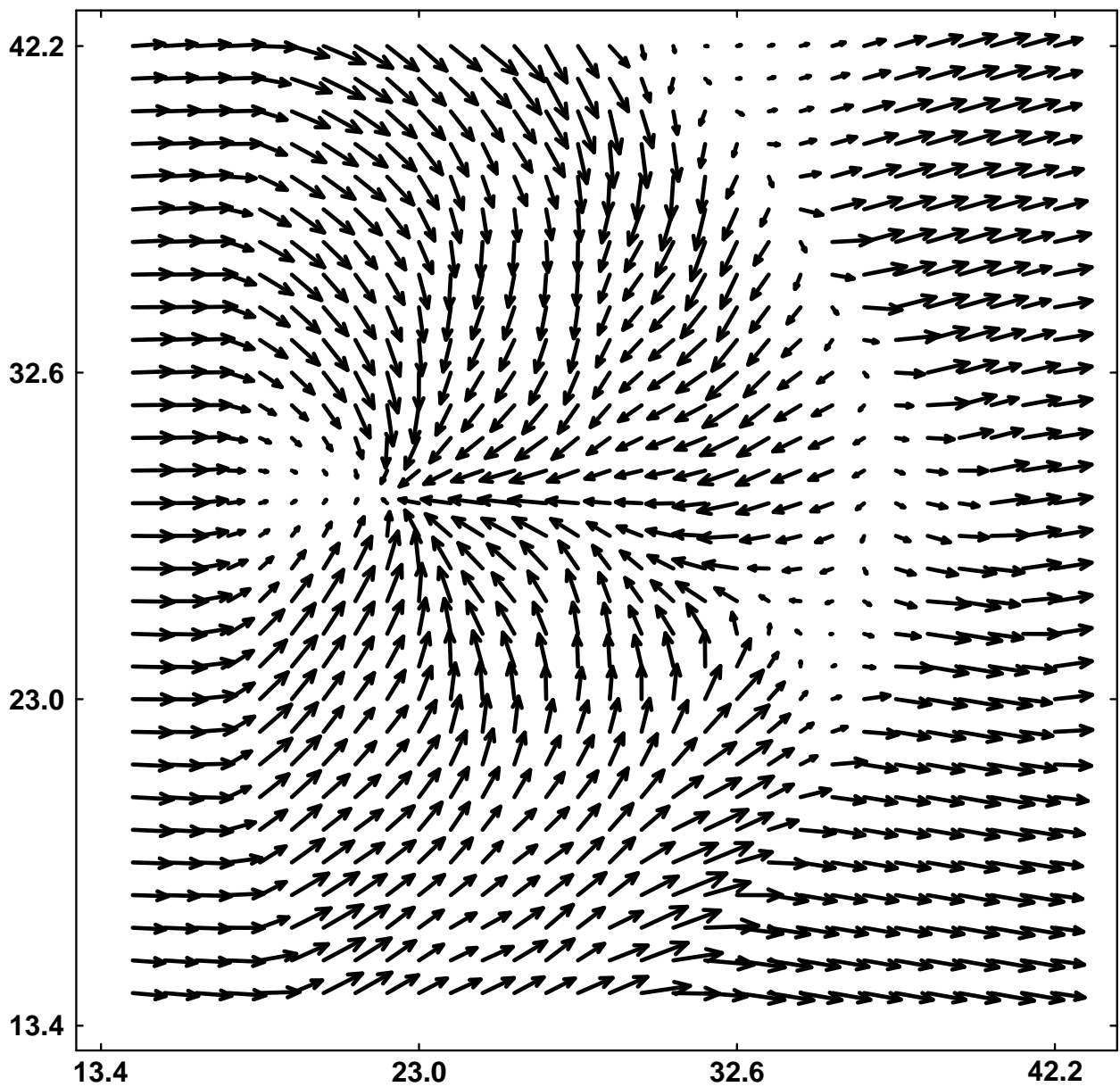
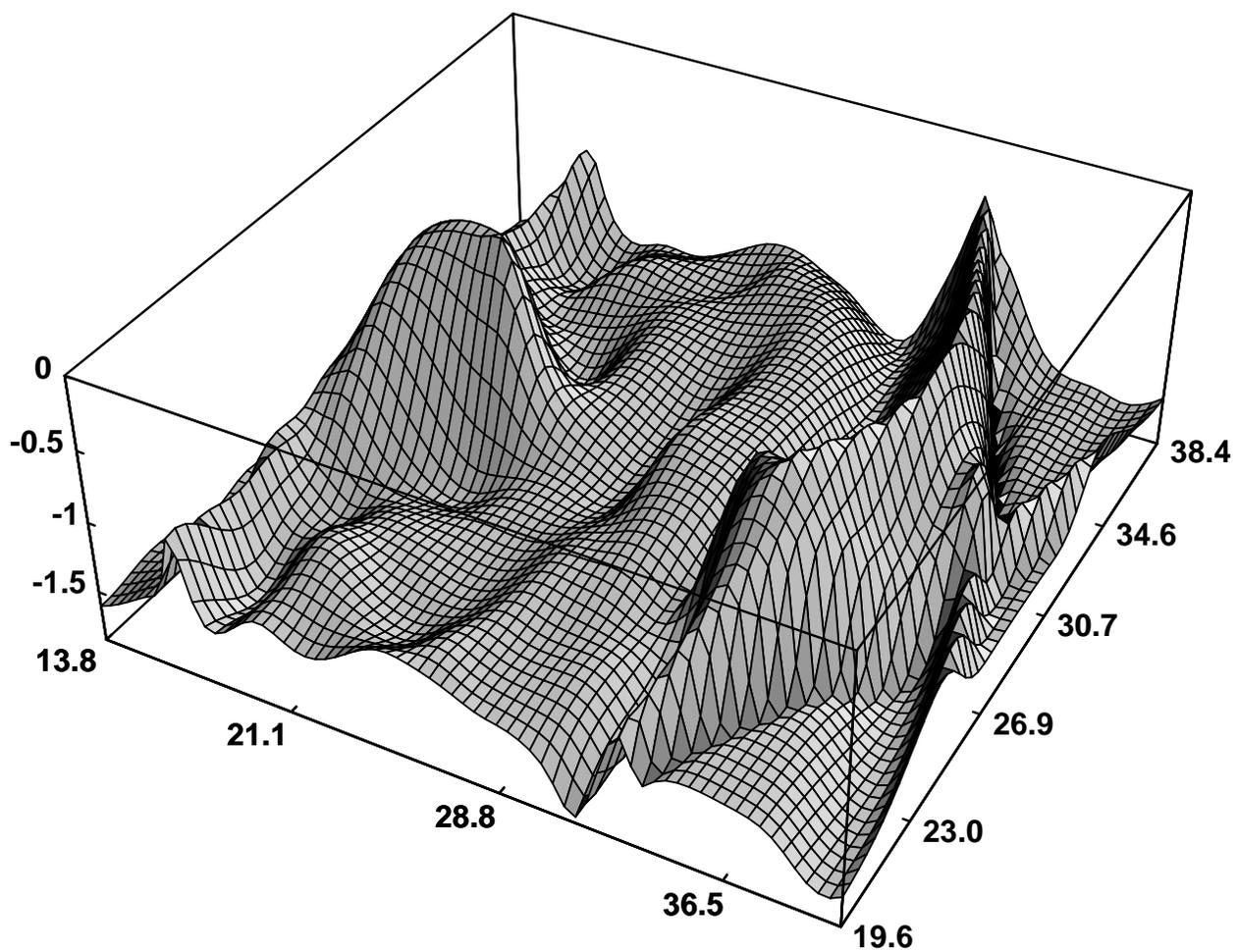
\end{document}